\documentclass[lettersize,journal]{IEEEtran}
\usepackage{newtxtext,newtxmath}
\usepackage{amsmath,amsfonts}
\usepackage{algorithmic}
\usepackage{algorithm}
\usepackage{array}
\usepackage{multirow}
\usepackage{bm}
\usepackage[caption=false,font=normalsize,labelfont=sf,textfont=sf]{subfig}
\usepackage{textcomp}
\usepackage{stfloats}
\usepackage{url}
\usepackage{verbatim}
\usepackage{cite}
\usepackage{booktabs}
\usepackage{multirow}
\usepackage{graphicx} % 用于 \resizebox
\usepackage{caption}
\usepackage{geometry}
\begin{document}

\title{SLIF-MR: Self-loop Iterative Fusion of Heterogeneous Auxiliary Information for Multimodal Recommendation}

\author{Jie Guo,~\IEEEmembership{Senior Member,~IEEE,}
Jiahao Jiang,
Ziyuan Guo,
Bin Song,~\IEEEmembership{Senior Member,~IEEE},
Yue Sun
        % <-this % stops a space
\thanks{Jie Guo, Jiahao Jiang, Ziyuan Guo and Bin Song are with the State Key Laboratory of Integrated Services Networks, Xidian University, Xi’an, Shaanxi 710071, China (e-mail: jguo@xidian.edu.cn; jiahaojiang@stu.xidian.edu.cn; ziyuanguo@stu.xidian.edu.cn; bsong@mail.xidian.edu.cn).\\ \hspace*{1em}Yue Sun is with the Health Science Center, School of Nursing, Xi'an Jiaotong University, Xi’an, Shaanxi 710049, China (e-mail: sun.yue@xjtu.edu.cn).}% <-this % stops a space
}

\markboth{Journal of \LaTeX\ Class Files,~Vol.~14, No.~8, August~2021}%
{Shell \MakeLowercase{\textit{et al.}}: A Sample Article Using IEEEtran.cls for IEEE Journals}


\maketitle

\begin{abstract}
Knowledge graphs (KGs) and multimodal item information, which respectively capture relational and attribute features, play a crucial role in improving recommender system accuracy. Recent studies have attempted to integrate them via  multimodal knowledge graphs (MKGs) to further enhance recommendation performance. However, existing methods typically freeze the MKG structure during training, which limits the full integration of structural information from heterogeneous graphs (e.g., KG and user-item interaction graph), and results in sub-optimal performance. To address this challenge, we propose a novel framework, termed \textbf{S}elf-\textbf{l}oop \textbf{I}terative \textbf{F}usion of Heterogeneous Auxiliary Information for \textbf{M}ultimodal \textbf{R}ecommendation (\textbf{SLIF-MR}), which leverages item representations from previous training epoch as feedback signals to dynamically optimize the heterogeneous graph structures composed of KG, multimodal item feature graph, and user-item interaction graph. Through this iterative fusion mechanism,  both user and item representations are refined, thus improving the final recommendation performance. Specifically, based on the feedback item representations, SLIF-MR constructs an item-item correlation graph, then integrated into the establishment process of heterogeneous graphs as additional new structural information in a self-loop manner. Consequently, the internal structures of heterogeneous graphs are updated with the feedback item representations during training. Moreover, a semantic consistency learning strategy is proposed to align heterogeneous item representations across modalities. The experimental results show that SLIF-MR significantly outperforms existing methods, particularly in terms of accuracy and robustness.
\end{abstract}


\section{Introduction}
\IEEEPARstart{I}{n} the era of mobile internet characterized by information overload, recommender systems have become essential on platforms such as short video apps, e-commerce sites, and news portals to deliver personalized services to users \cite{shang2023learning,wang2023industrial}. Traditional recommendation methods, including matrix factorization and collaborative filtering \cite{wang2019neural,he2020lightgcn}, rely primarily on user-item interaction data to predict user preferences. However, they encounter significant challenges in addressing data sparsity and the cold-start problem.

\begin{figure}[t]
  \centering
  \includegraphics[width=\linewidth]{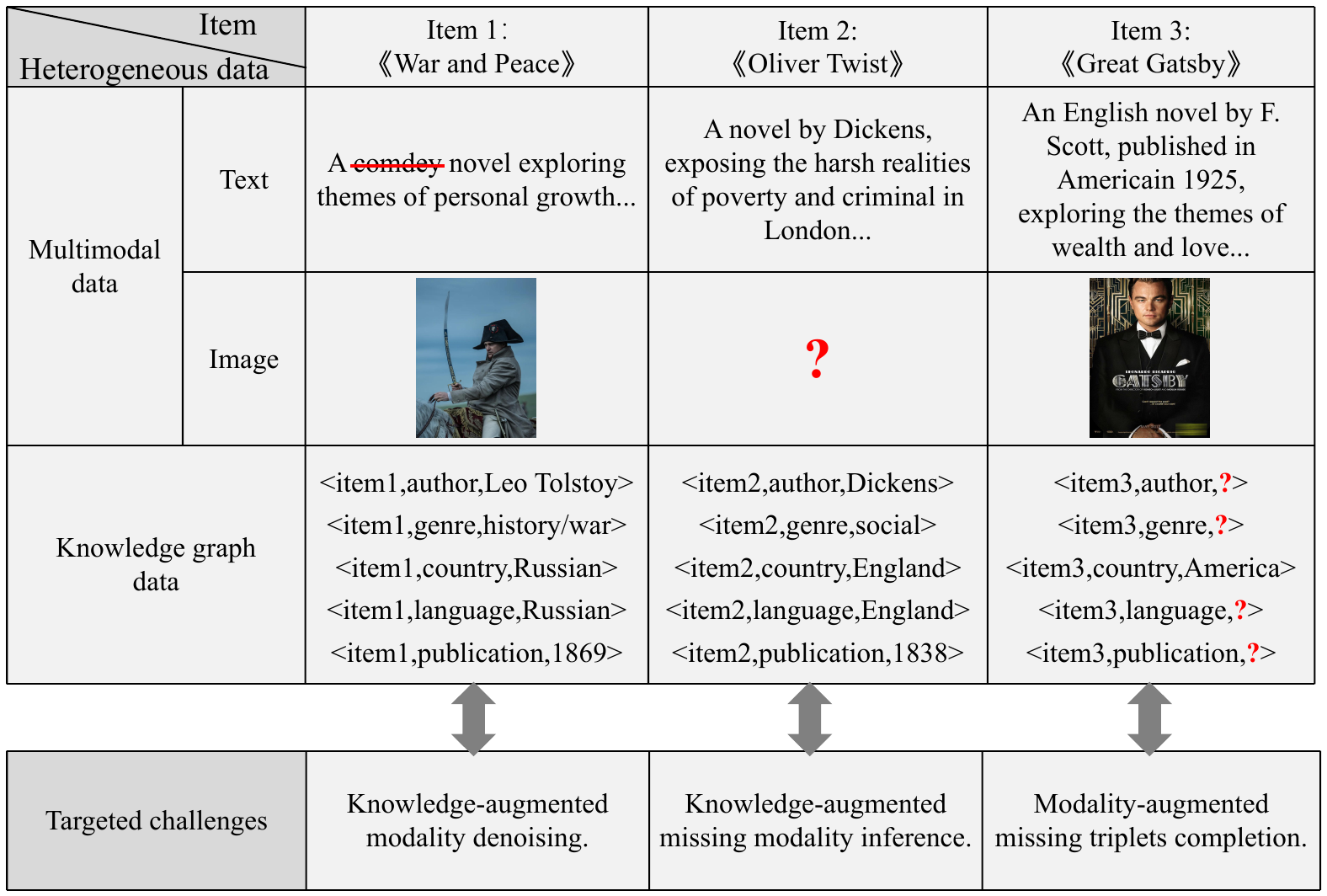} 
  % \vspace{-12pt}
  \caption{Illustration of deficiencies in KG and multimodal data for recommendation. In this figure, the symbol $``-"$ denotes noise data, and the symbol $``?"$ indicates missing data.}
  \label{Fig.1}
\end{figure}

In recent years, researchers have introduced diverse forms of auxiliary information (e.g.,  knowledge graph, multimodal item information), injecting new vitality into the community of recommender systems, including Multimodal Recommender Systems (MRSs) \cite{liu2024inter, liu2023multimodal, guo2024space} and Knowledge Graph-based Recommender Systems (KGRs) \cite{anand2024survey}, which have significantly improved recommendation accuracy. Nevertheless, methods relying on a single type of auxiliary information inherently suffer from notable limitations. Specifically, KGRs face the challenge of knowledge imbalance \cite{tang2024editkg,zhang2024relation,yu2024knowledge}, where popular items possess substantially more attributes than cold-start items. As illustrated in Figure \ref{Fig.1}, item 3, \textit{``Great Gatsby"} lacks several critical attributes, which may lead to its marginalization during model training and result in biased recommendation. On the other hand, MRSs are plagued by issues of modality noise and modality absence \cite{malitesta2024we,xv2024improving}. For instance, item 1, \textit{``War and Peace"} is inaccurately classified as a ``comedy novel", while item 2, \textit{``Oliver Twist"} lacks corresponding image data. These challenges severely limit the performance of MRSs \cite{liu2024multimodal}.

To address these issues, some researchers have integrated the knowledge graph with multimodal item data through the multimodal knowledge graph (MKG) to further enhance recommendation performance \cite{sun2020multi,lin2023automatic}. However, we argue that these methods exhibit two significant limitations: (i) Entity Heterogeneity. There exists a notable discrepancy between the representations of image entities extracted from pre-trained ResNet model and text entities extracted from BERT model, which impedes the seamless integration of the two modalities. (ii) Relationship Heterogeneity. For instance, a book knowledge graph explicitly represent semantic correlations between book entities and their attributes through fine-grained relationships such as \textit{``type,"} \textit{``author,"} and \textit{``language."} In contrast, current MKG typically employs coarse-grained relationships, such as using the \textit{``textual description is"} relationship to link book entities with their textual descriptions, while overlooking the fact that textual descriptions may encompass multiple attributes. The heterogeneity of entities and relationships in MKG restricts the recommender system's ability to effectively integrate multimodal data with the knowledge graph, potentially leading to the loss of critical information during model training and ultimately decreasing recommendation accuracy.

To address the aforementioned challenges, we propose a novel framework, termed Self-loop Iterative Fusion of Heterogeneous Auxiliary Information for Multimodal Recommendation (SLIF-MR). The proposed framework leverages item representations from previous training epoch as feedback signals to dynamically optimize heterogeneous graph structures composed of the knowledge graph (KG), multimodal item data and user-item interaction data, while incorporating semantic consistency learning to refine both user and item representations. Specifically, SLIF-MR applies graph neural networks (GNNs) to propagate and aggregate information across the KG, the item feature graph, and the user-item interaction graph to learn heterogeneous item representations. These representations are then weighted and aggregated via an item-level attention mechanism to generate unified item representations that encapsulate rich semantic information from heterogeneous graphs. Then, we construct an item-item correlation graph using the unified item representations and inject it as new structural information into the heterogeneous graphs in a self-loop manner. During the model training process, the structures of the heterogeneous graphs are continuously updated to progressively align with real user preferences and item correlations. Furthermore, we introduce a semantic consistency learning method to further unify the heterogeneous item representations by aligning them within the shared vector space.
Our primary contributions can be outlined as follows:
\begin{itemize}
    \item We propose a novel self-loop iterative fusion method for sufficient integration of heterogeneous auxiliary information. It adopts a dynamic optimization mechanism, which utilizes item representations from the previous training epoch to update the internal graph structures composed of KG, multimodal item data and user-item interaction data.
    \item We propose the semantic consistency learning method to overcome the semantic representation difference from the perspectives of item multimodal semantic relation and user-item occurrence relation and ultimately refine both user and item representations.
    \item The proposed SLIF-MR model exhibits superior performance compared to state-of-the-art Multimodal Recommender Systems (MRSs) and Knowledge Graph-based Recommender Systems (KGRs), validating the effectiveness of our approach.
\end{itemize}

\section{PRELIMINARIES}
We begin by introducing structured data and key concepts relevant to our investigation and then formally define the heterogeneous information-based recommendation task.

\textbf{User-Item Interaction Graph.}
In this paper, we model the users' implicit feedback (e.g., purchases, clicks, etc.) as a bipartite graph $\mathcal{G}_b = \left\{ (u, i,y_{u, i}) | u \in \mathcal{U}, i \in \mathcal{I} \right\}$, where $\mathcal{U}$ and $\mathcal{I}$ represent the user set and item set, respectively, and $ |\mathcal{U}| $ and $ |\mathcal{I}| $ represent the number of users and items, respectively. If an interaction exists between user $u$ and item $i$, $y_{u,i} = 1$; otherwise, $y_{u,i} = 0$.

\textbf{Knowledge Graph.}
The knowledge graph stores various attributes associated with items in the form of triples. We formally represent it as $\mathcal{G}_k = \left\{ (h,r,t) | h,t \in \mathcal{E}, r \in \mathcal{R} \right\}$, where $\mathcal{E}$ and $\mathcal{R}$ represent the entity set and relation set, respectively. Specifically, the head entity $h$ and the tail entity $t$ are connected by the relation $r$. For instance, the triple (\textit{The Fairy Tale King, \  Book.Genre,  \ Children}) indicates that \textit{``The Fairy Tale King"} is a children's book. In our study, the item set $\mathcal{I}$ is a subset of the entity set $\mathcal{E}$.

\textbf{Multimodal Item Features.}
In this study, we consider the modalities $\mathcal{M}=\{\mathcal{V},\mathcal{T}\}$, where $\mathcal{V}$ and $\mathcal{T}$ denote the visual and the textual modality respectively. Furthermore, the proposed SLIF-MR framework can be extended to incorporate additional modalities.

\section{METHODOLOGY}
% We now introduce the proposed SLIF-MR framework in detail. As illustrated in Figure \ref{Fig:2}, SLIF-MR consists of three core modules: (1) Heterogeneous representation extraction module. We first adopt mainstream methods to construct the item feature graph based on multimodal item features \cite{zhou2023tale, zhou2023attention}, then apply GNN to extract heterogeneous item representations from the item feature graph, knowledge graph, and interaction graph, respectively. (2) Heterogeneous self-loop graph enhancement module. We aggregate heterogeneous item representations via an item-level attention mechanism to generate unified item representations that contain rich semantic information. Then, we construct an item correlation graph by calculating pairwise similarities between item representations and inject this graph as new structural information into the heterogeneous graph structures in a self-loop manner. (3) Heterogeneous Representations Consistency Fusion Module. We propose a semantic consistency fusion method to further unify heterogeneous item representations by aligning them within a shared vector space.

\begin{figure*}[t]
  \centering
  \includegraphics[width=\textwidth]{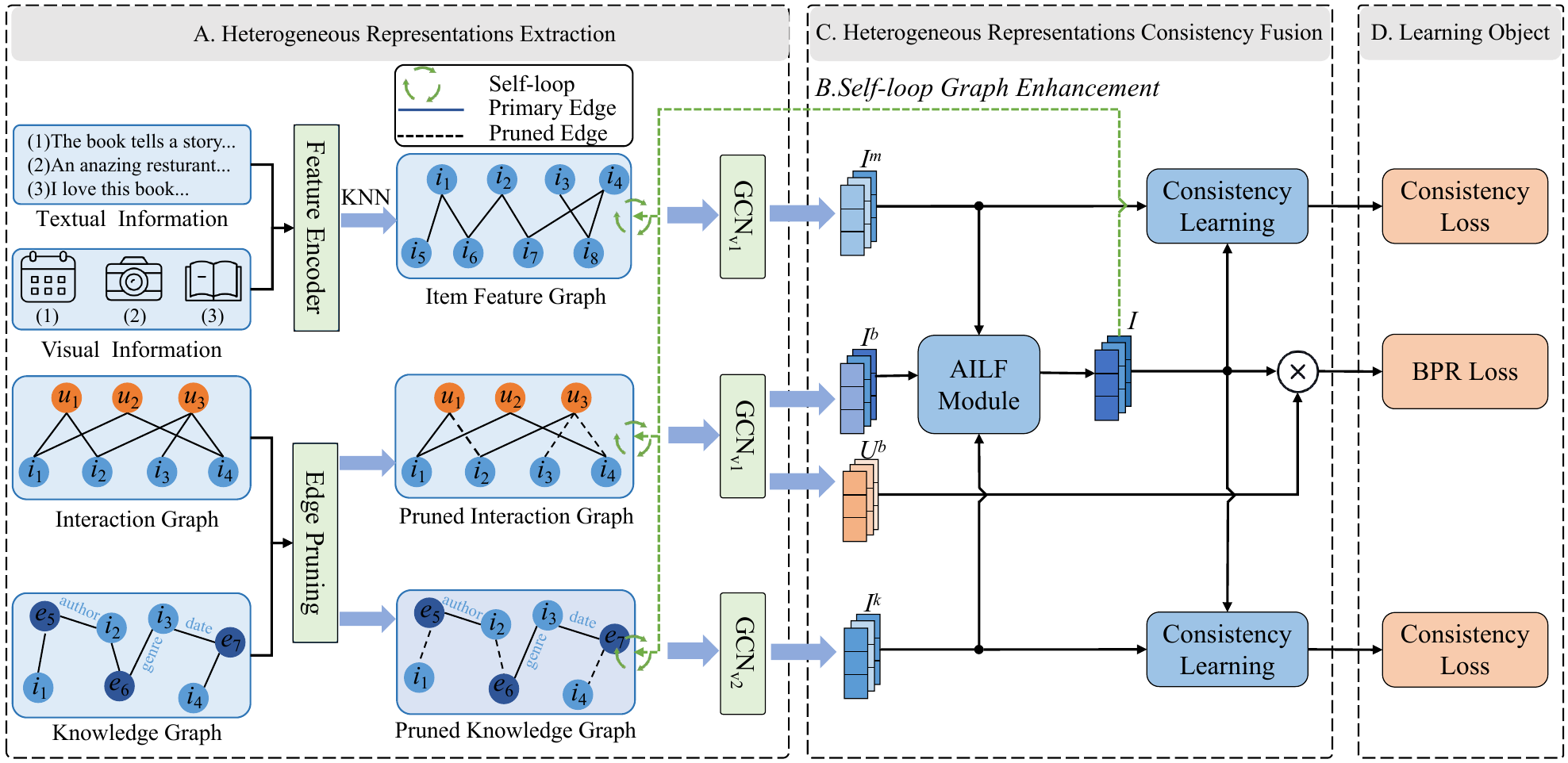}
   % \vspace{-15pt}
  \caption{Framework of the proposed SLIF-MR model.}
  \label{Fig:2}
\end{figure*}

We now provide a detailed description of the proposed SLIF-MR framework. As shown in Figure \ref{Fig:2}, SLIF-MR comprises three key modules: (i) Heterogeneous Representation Extraction Module: In this module, we first construct the item feature graph based on multimodal item features. Subsequently, we employ Graph Neural Networks (GNNs) to extract heterogeneous item representations from the item feature graph, knowledge graph, and interaction graph. (ii) Self-loop Heterogeneous Graph Enhancement Module: Here, we aggregate heterogeneous item representations using an item-level attention mechanism to generate unified item representations that capture rich semantic information. We then build an item correlation graph by calculating pairwise similarities between item representations, which is subsequently injected as new structural information into the heterogeneous graph structures in a self-loop manner. (iii) Heterogeneous Representations Consistency Fusion Module: In this final module, we introduce a semantic consistency fusion method to further harmonize heterogeneous item representations by aligning them within a shared vector space.

\subsection{Heterogeneous Representations Extraction Module}

\subsubsection{Graph Construction}
We first construct the interaction graph adjacency matrix $\bm{A}$ based on user-item interaction data:
\begin{equation}
\bm{A} =
\begin{bmatrix}
\bm{0} & \bm{R} \\
\bm{R}^\top & \bm{0}
\end{bmatrix},
\end{equation}
where \( \bm{R} \in \mathbb{R}^{|\mathcal{U}| \times |\mathcal{I}|} \) denotes the user-item interaction matrix.

To further exploit latent item correlations embedded in the unified representations, we construct an item correlation graph $\bm{S}$ by calculating pairwise similarities $\hat{\bm{S}}_{ij}$ between item representations.

\begin{equation}
    \hat{{S}}_{ij}^{m} = \frac{(x_i^{m})^\top x_j^{m}}{\lVert x_i^{m}\rVert \cdot \lVert x_j^{m}\rVert},
\end{equation}
where $m \in \{\mathcal{V}, \mathcal{T}\}$, and $x_i^{m}$ represents the raw feature of item $i$ in modality $m$. Then, we remove noisy edges by performing k-nearest neighbors (KNN) sparsification operation on $\hat{\bm{S}}^{m}$, retaining the top-$K$ edges with the highest confidence scores for each item:

\begin{equation}
\tilde{{S}}_{ij}^m = 
\begin{cases}
\hat{{S}}_{ij}^m, & \text{if } \hat{{S}}_{ij}^m \in \operatorname{top-K}(\hat{\bm{S}}_i^m), \\
0, & \text{otherwise}.
\end{cases}
\end{equation}

Furthermore, to mitigate the risk of gradient vanishing and explosion, we apply the Laplace Normalization operation to $\widetilde{\bm{S}}^m$:

\begin{equation}
\bm{S}^m = (\bm{D}^m)^{-\frac{1}{2}} \tilde{\bm{S}}^m (\bm{D}^m)^{-\frac{1}{2}},
\end{equation}
where $\bm{D}^m$ denotes the diagonal degree matrix of $\tilde{\bm{S}}^m$. 

Finally, the item feature graph $\bm{S} \in \mathbb{R}^{|\mathcal{I}| \times |\mathcal{I}|}$ can be constructed by combining the normalized matrix of each modality:

\begin{equation}
    \bm{S} = \sum_{m \in \mathcal{M}} \alpha_m \bm{S}^m,
\end{equation}
where \( \alpha_m \) denotes the weight coefficient of modality $m$, which can be automatically adjusted by backpropagation during model training.

\subsubsection{Graph Edge Pruning}
To mitigate the long-tail problem in the knowledge graph $\mathcal{G}_k$ and user-item interaction graph $\bm{A}$, we perform degree-sensitive edge pruning strategy on them. Specifically, we pre-compute the probability $p_e$ that each edge in the graph structure is preserved:
\begin{equation}
    p_e = \frac{1}{\sqrt{d_{m}} \sqrt{d_{n}}},
\end{equation}
where $m$ and $n$ represent the two endpoints of edge $e$, $d_m$ and $d_n$ are the degrees of the two endpoints (i.e., the number of nodes connected to them in the graph). Therefore, edges connected by popular nodes have a higher probability of being dropped.
In addition, we control the number of droped edges by setting the pruning ratio.

\subsubsection{Information Propagation and Aggregation}
In this subsection, we employ the graph neural networks to derive heterogeneous item representations from the user-item interaction graph $\bm{A}$, item feature graph $\bm{S}$, and knowledge graph $\mathcal{G}_k$, respectively.

Following mainstream recommendation methods \cite{he2020lightgcn,wang2019kgat,wang2019neural}, we initialize the user and item representations using ID embeddings $x^u, x^i \in \mathbb{R}^{d}$, respectively, where $d$ represents the embedding dimension. Taking the initialized ID embeddings of users and items as the input, we first perform the lightweight graph convolution operation \cite{he2020lightgcn} on interaction graph $\bm{A}$. The representations at $l$-th layer can be expressed as:

\begin{equation}
\begin{aligned}
\bar{x}_u^l &= \sum_{i \in N_u} \frac{1}{\sqrt{|N_u|}\sqrt{|N_i|}} \bar{x}_i^{l-1}, \\
\bar{x}_i^l &= \sum_{u \in N_i} \frac{1}{\sqrt{|N_i|}\sqrt{|N_u|}} \bar{x}_u^{l-1},
\end{aligned}
\end{equation}
where $N(u)$ is the set of items interacting with user $u$, and $N(i)$ is the set of usres interacting with item $i$. The  symmetric normalization term $\tfrac{1}{\sqrt{|N_i|}\sqrt{|N_u|}}$ is utilized to avoid the scale of features increasing with graph convolution operation.

Furthermore, we stack multiple convolutional layers and aggregate their outputs via average pooling operation, resulting in the final representations of user $u$ and item $i$ derived from the interaction graph $\bm{A}$:

\begin{equation}
\begin{aligned}
\bar{x}_u &= \frac{1}{L_{b}} \sum_{l=1}^{L_{b}} \bar{x}_u^{l} + \bar{x}_u^0,\\
\bar{x}_i &= \frac{1}{L_{b}} \sum_{l=1}^{L_{b}} \bar{x}_i^{l} + \bar{x}_i^0,
\end{aligned}
\end{equation}
where $L_{b}$ denotes the number of convolutional layers. And to mitigate the issue of excessive smoothing during graph convolution \cite{mao2021ultragcn}, we employ a residual linkage approach by incorporating the initial representation $\bar{x}_u^0$ and $\bar{x}_i^0$ (i.e., original user and item ID embeddings).

Subsequently, we perform similar graph convolution operation on the item feature graph $\bm{S}$:

\begin{equation}
\begin{aligned}
\tilde{x}_i^l &= \sum_{\substack{j \in N(i) \\ i \neq j}} \bm{S}_{ij} \tilde{x}_j^{l-1},\\
\tilde{x}_i &= \frac{1}{L_{m}} \sum_{l=1}^{L_{m}} \tilde{x}_i^{l} + \tilde{x}_i^0,
\end{aligned}
\end{equation}
where $\tilde{x}_i$ denotes the representation of item $i$ extracted from item feature graph $\bm{S}$, and $L_m$ is the number of convolution layers for $\bm{S}$.

For information propagation and aggregation on the knowledge graph $\mathcal{G}_k$, it is crucial to consider the diverse relationships between entities, as these relationships reflect complex semantic correlations among entities (including items). Given a head entity $h$, the graph convolution layer on $\mathcal{G}_k$ is defined as follows:

\begin{equation}
\hat{{x}}_{h}^{l} = \sum_{(h,r,t)\in N_h}\alpha(h,r,t)\hat{{x}}_t^{l-1},
\end{equation}
where $N_h \in \{(h, r, t)| (h, r, t) \in \mathcal{G}_k\}$ denotes the set of triples in which $h$ serves as the head entity. The term $\alpha(h,r,t)$ represents the attenuation factor for the propagation of information, which controls the extent to which information is transmitted from the tail entity $t$ to head entity $h$ via the relation $r$. $\alpha(h,r,t)$ is calculated as follows:

\begin{equation}
\hat{\alpha}(h, r, t) = (\bm{W}_r \hat{x}_t)^\top \text{tanh}\left(\bm{W}_r \hat{x}_h + \hat{x}_r\right),
\end{equation}
where $\bm{W}_r \in \mathbb{R}^{d \times d}$ denotes the transformation matrix for relation $r$, used to project  head entity $h$ and tail entity $t$ into the space of relation $r$. $\text{tanh}(\cdot)$ represents the nonlinear activation function, which ensures that the attenuation factor depends on the distance between $h$ and $t$ in the space of relation $r$. 

Then, we use the softmax function to normalize the attenuation factors among all triples connected by $h$:

\begin{equation}
\alpha(h, r, t) = \frac{\exp(\hat{\alpha}(h, r, t))}{\sum_{(h, r', t') \in N_h} \exp(\hat{\alpha}(h, r', t'))}.
\end{equation}

Finally, we calculate the representation of entity $h$ following the method described in Equation (8):

\begin{equation}
    \hat{x}_h = \frac{1}{L_{k}+1} \sum_{l=0}^{L_{k}} \hat{x}_h^{l},
\end{equation}
where $L_{k}$ denotes the number of convolutional layers for knowledge graph.

\subsection{Self-loop Heterogeneous Graphs Enhancement Module}
In this section, we aggregate the heterogeneous item representations derived from the previous subsection to generate unified item representations. These representations serve as feedback signals, guiding the optimization of upstream heterogeneous graph structures in a self-loop manner, allowing them to gradually align with true user preferences and item correlations.

\begin{figure}[t]
  \centering
  \includegraphics[width=0.9\columnwidth]{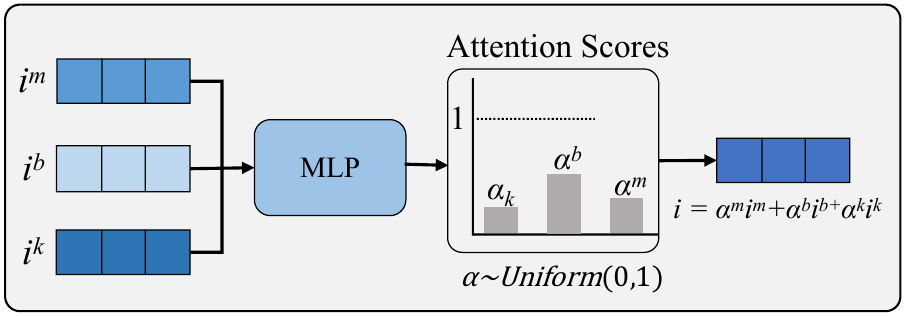}
  % \vspace{-5pt}
  \caption{AILF: attention-based item-level fusion.}
  \label{Fig:3}
\end{figure}

\subsubsection{Attention-Based Item-Level Fusion} 
Considering the varying contributions of heterogeneous information to the representations of different items, we employ an item-level attention mechanism to learn personalized weightings for each item’s heterogeneous representation. As illustrated in Figure \ref{Fig:3}, we first input the heterogeneous item representations into the attention module to obtain the weight vector $\bm{\alpha}_i = (\alpha_i^b,\alpha_i^m,\alpha_i^k)$:

\begin{equation}
\bm{\alpha}_i = \text{Softmax}\left(\frac{\bm{W} \cdot [\bar{x}_i |.| \tilde{x}_i |.| \hat{x}_i] + \bm{b}}{\sqrt{d}}\right),
\end{equation}
where $\bm{W}\in\mathbb{R}^{3d \times 3}$ is a learnable transformation matrix, used to map heterogeneous item representations to a uniform vector space, and $\mathbf{b}\in \mathbb{R}^{3}$ is a learnable bias vector. The symbol $|.|$ represents the concatenation operation, and ${\sqrt{d}}$ is a scaling factor applied to the attention weight $\bm{\alpha}_i$ to stabilize the variance of the dot product results, thereby mitigating gradient instability.

Next, we perform weighted aggregation of heterogeneous item representations to obtain the unified item representation ${x}_i$:

\begin{equation}
    {x}_i = \alpha_i^{b} \bar{x}_i\ + \alpha_i^{m} \tilde{x}_i\ + \alpha_i^{k} \hat{x}_i.
\end{equation}

\subsubsection{Self-loop Heterogeneous Graphs Enhancement}
\begin{figure}[t]
  \centering
  \includegraphics[width=\columnwidth]{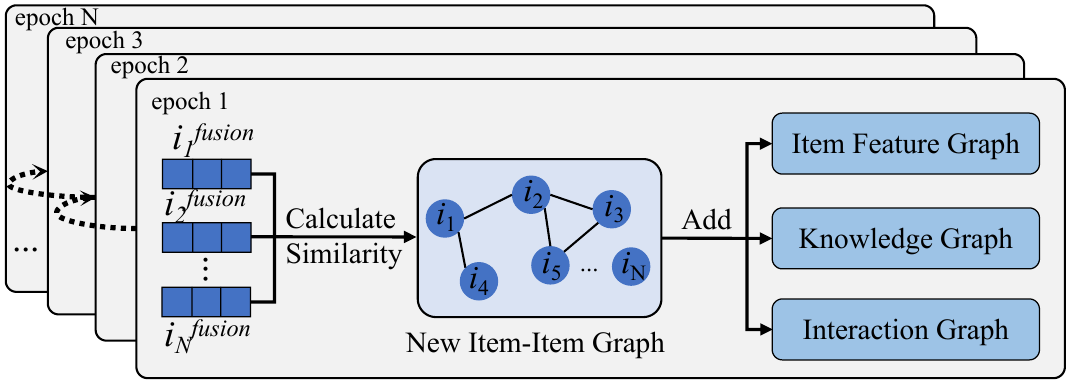}
  % \vspace{-13pt}
  \caption{Self-loop heterogeneous graphs enhancement.}
  \label{Fig:4}
\end{figure}

The latent item-item correlations inherent in interaction data and auxiliary information have been shown to enhance the expressive power of graph models, significantly improving recommendation performance \cite{zhou2023tale,zhou2023attention,zhang2021mining}. In the user-item interaction graph, this latent correlation is characterized as \textit{``items interacted by the same user may be similar"}; while in the knowledge graph, it is articulated as \textit{``items sharing identical attributes or relationships may be similar"}. By leveraging various types of auxiliary information, the latent correlations between items become more pronounced and multidimensional. Consequently, we propose to model the latent item-item correlation embedded within heterogeneous auxiliary information to uncover potential user preferences. As illustrated in Figure \ref{Fig:4}, our methodology commences with the computation of pairwise similarities between unified item representations:

\begin{equation}
    \hat{{S}}_{ij} = \frac{(x_i)^\top x_j}{\lVert x_i \rVert \cdot \lVert x_j \rVert}.
\end{equation}

Subsequently, we sequentially apply top-$N$ sparsification and normalization operations to $\hat{\bm{S}}$, constructing an item-item correlation graph denoted as $\mathcal{G}$:
\begin{equation}
\begin{aligned}
\tilde{\mathcal{G}}_{ij} &= 
\begin{cases}
\hat{\bm{S}}_{ij}, & \text{if } \hat{\bm{S}}_{ij} \in \operatorname{top-N}(\hat{\bm{S}}_{i}), \\
0, & \text{otherwise},
\end{cases} \\
\mathcal{G} &= (\bm{D})^{-\frac{1}{2}} \tilde{\mathcal{G}} (\bm{D})^{-\frac{1}{2}},
\end{aligned}
\end{equation}
where $\bm{D}$ denotes the diagonal degree matrix of $\tilde{\mathcal{G}}$.

Next, we incorporate the item-item correlation graph $\mathcal{G}$ into the heterogeneous graph structures as additional new structural information, including the interaction graph $\bm{A}$, knowledge graph $\mathcal{G}_k$, and item feature graph $\bm{S}$. Furthermore, we execute the aforementioned process in a self-loop manner to dynamically refine item-item correlations during model training, thereby enabling the heterogeneous graph structures to progressively converge with real user preferences:

\begin{equation}
\begin{aligned}
\bm{A}^n &= 
\begin{bmatrix}
\bm{0} & \bm{R} \\
\bm{R}^\top & \mathcal{G}^{n}
\end{bmatrix},\\
\bm{S}^n &= [\bm{S}^{n-1}\ |.|\ \mathcal{G}^{n}],\\
\mathcal{G}_k^{n} &= [\mathcal{G}_k^{n-1} |.| \mathcal{G}^{n}],
\end{aligned}
\end{equation}
where $\mathcal{G}^{n}$ denotes the item-item correlation graph constructed in the $n$-th epoch of model training, and $\bm{A}^n$, $\bm{S}^n$ and $\mathcal{G}_k^{n}$ represent the $n$-th enhanced interaction graph, item feature graph and knowledge graph, respectively. 

\subsection{Heterogeneous Representations Consistency Fusion}
Significant semantic discrepancies exist across heterogeneous data, including interaction data, knowledge graph, and multimodal item data. As a result, the item representations derived from these heterogeneous sources exhibit inherent semantic heterogeneity. In this subsection, we aim to obtain precise item representations through the proposed semantic consistency learning strategy. For conciseness, we will refer to the heterogeneous item representations discussed in subsection A as distinct modalities in the following discussion.

\subsubsection{Inter-Modal Semantic Consistency}
To address cross-modal semantic heterogeneity challenge and improve the fusion of heterogeneous item representations, we introduce the inter-modal semantic consistency loss. Specifically, we adopt the unified item representation derived in Equation (14) as a benchmark and aim to unidirectionally align the item representations across different modalities with this benchmark. Formally:

\begin{equation}
\begin{aligned}
\mathcal{L}_{\text{inter}}^{B}&= - \sum_{i \in I} \log \frac{\exp(\operatorname{sim}(\bar{x}_i, sg(x_i)) / \tau)}{\sum_{j \in I \setminus \{i\}} \exp(\operatorname{sim}(\bar{x}_i, sg(x_j))) / \tau)},\\
\mathcal{L}_{\text{inter}}^{K}&= - \sum_{i \in I} \log \frac{\exp(\operatorname{sim}(\hat{x}_i, sg(x_i)) / \tau)}{\sum_{j \in I \setminus \{i\}} \exp(\operatorname{sim}(\hat{x}_i, sg(x_j))) / \tau)},\\
\mathcal{L}_{\text{inter}}^{M}&= - \sum_{i \in I} \log \frac{\exp(\operatorname{sim}(\tilde{x}_i, sg(x_i)) / \tau)}{\sum_{j \in I \setminus \{i\}} \exp(\operatorname{sim}(\tilde{x}_i, sg(x_j)) / \tau)},
\end{aligned}
\end{equation}
where $x_i$ denotes the unified representation of item $i$ from subsection 3.2, and $\bar{x}_i$, $\hat{x}_i$ and $\tilde{x}_i$ are heterogeneous representations extracted from interaction graph, knowledge graph and item feature graph, respectively. The notation $\text{sim}(\cdot)$ denotes cosine similarity, and $\tau$ represents the temperature parameter. 

Additionally, the notation $sg(\cdot)$ represents the stop gradient backward operation, which ensures unidirectional alignment and minimizes noise modality interference as much as possible \cite{li2024align}. The total inter-modal consistency loss $\mathcal{L}_{\text{inter}}$ is defined as:

\begin{equation}
    \mathcal{L}_{\text{inter}}= \mathcal{L}_{\text{inter}}^{B} + \mathcal{L}_{\text{inter}}^{K} + \mathcal{L}_{\text{inter}}^{M} 
\end{equation}

\subsubsection{Intra-modal Semantic Consistency}
It is crucial to emphasize that the core concept of cross-modal semantic consistency loss lies in reducing semantic discrepancies by aligning heterogeneous representations of the same item in the vector space. Although effective, this approach introduces, to some extent, the issue of intra-modal semantic distortion. Specifically, within a single modality, representations of dissimilar items may be positioned closer together, while those of similar items may be pushed further apart, resulting in the misalignment of intra-modal semantic correlations. To tackle this challenge, we introduce a novel methodology of constructing intra-modal semantic consistency loss:

\begin{equation}
\begin{aligned}
\mathcal{L}_{\text{intra}}^{B} &= -\frac{1}{|I|^2} \sum_{i, j=1}^{I} \log \left( e^{-t \| f(\bar{x}_i) - f(\bar{x}_j) \|^2} \right),\\
\mathcal{L}_{\text{intra}}^{K} &= -\frac{1}{|I|^2} \sum_{i, j=1}^{I} \log \left( e^{-t \| f(\hat{x}_i) - f(\hat{x}_j) \|^2} \right),\\
\mathcal{L}_{\text{intra}}^{M} &= -\frac{1}{|I|^2} \sum_{i, j=1}^{I} \log \left( e^{-t \| f(\tilde{x}_i) - f(\tilde{x}_j) \|^2} \right),
\end{aligned}
\end{equation}
where $t$ denotes the temperature parameter, and $ \| f(\mathbf{x}_i) - f(\mathbf{x}_j) \|^2$ represents the squared Euclidean distance between the embeddings of item $i$ and item $j$. The core principle of intra-modal consistency loss is to maintain the semantic consistency of items within a modality by minimizing the mutual information between item representations in the same modality and promoting a relatively uniform distribution of distances between them. The total intra-modal consistency loss $\mathcal{L}_{\text{intra}}$ is defined as:

\begin{equation}
    \mathcal{L}_{\text{intra}}= \mathcal{L}_{\text{intra}}^{B} + \mathcal{L}_{\text{intra}}^{K} + \mathcal{L}_{\text{intra}}^{M}
\end{equation}

\subsection{Optimization Object}
We employ the Bayesian Personalised Ranking (BPR) loss to optimize the proposed model, encouraging higher prediction scores for observed interactions than unobserved interactions:

\begin{equation}
    \mathcal{L}_{\text{bpr}} = - \sum_{u \in U} \sum_{i \in N_u} \sum_{j \notin N_u} \ln \sigma(y_{ui} - y_{uj}),
\end{equation}
where $N_u$ denotes the set of items that user $u$ has interacted with, $j \notin N_u$ represents the negative samples sampled from the unobserved interactions, $y_{ui}$ and $y_{uj}$ represent the predicted scores for positive and negative samples, respectively, and $\sigma(\cdot)$ represents the sigmoid function. Finally, we can get the learning object: 

\begin{equation}
    \mathcal{L} = \mathcal{L}_{\text{bpr}} + \beta\mathcal{L}_{\text{inter}} + \gamma\mathcal{L}_{\text{intra}} + \eta \lVert \Theta \rVert^2,
\end{equation}
where $\beta$ and $\gamma$ are the weighting coefficients of inter-modal consistency loss and intra-modal consistency loss, respectively. $\lVert \Theta \rVert^2$ denotes the regularization term and $\eta$ is its coefficient.

\section{EXPERIMENTS}
We present empirical results that demonstrate the validity of our proposed SLIF-MR framework. The experiment aims to address the following four research questions:

\begin{itemize}
\item {\textbf{RQ1}}: How does the performance of the proposed SLIF-MR framework compare to state-of-the-art baseline models?
\item {\textbf{RQ2}}: How do key components of the proposed SLIF-MR framework influence the overall model performance?
\item {\textbf{RQ3}}: How do hyper-parameters affect the performance of SLIF-MR?
\item {\textbf{RQ4}}: To what extent does the proposed SLIF-MR framework demonstrate robustness when subjected to various types of noise?
\item {\textbf{RQ5}}: How does SLIF-MR balance the trade-off between recommendation performance and computational complexity by adjusting the training epoch interval for heterogeneous graph updates?
\end{itemize}

\subsection{Experimental Settings}
\subsubsection{Datasets Description}
We conduct experiments on two benchmark datasets from different domains: Amazon-Book and Yelp2018.

\begin{itemize}
\item {\textbf{Amazon-book}}. 
This dataset is a large-scale book review collection provided by Amazon company, encompassing real user purchase records, corresponding textual descriptions and book images. In our study, we utilize the 2014 version of this dataset.
\item {\textbf{Yelp2018}}. 
This dataset is a merchant rating collection provided by Yelp, a well-known review platform, comprising real merchant rating records from multiple cities, along with merchant images and textual descriptions. In our study, we utilize the 2018 version of this dataset.
\end{itemize}
Building on prior studies \cite{zhao2019kb4rec,wang2019kgat}, we map items to Freebase entities and gather two-hop neighboring entities to construct the knowledge graph for each dataset. Both datasets encompass item information in textual and visual modalities. For the Amazon-book dataset, we leverage pre-extracted and publicly available 4096-dimensional visual features. Text embeddings are generated by concatenating the title, description, category, and brand of each item, followed by applying the BERT model \cite{reimers2019sentence} to obtain 768-dimensional text embeddings. For the Yelp dataset, we utilize the ResNet50 model \cite{he2016deep} to extract visual features from pre-collected images of each business. Text embeddings are generated by concatenating the name, region, star rating, and category information of each business, and then extracting them using the same BERT model \cite{reimers2019sentence}. For interaction data, we apply a 10-core filtering approach to retain users and items with at least 10 interactions. For each dataset, we allocate 80\% of each user's interaction records to the training set, 10\% to the validation set, and the remaining 10\% to the test set.

\begin{table}[t]
\centering  % 添加居中命令
\caption{Statistics of datasets.}
% \vspace{-10pt} % 调整为合适的值
\begin{tabular}{l|cc}
\hline
Stats. & Amazon-Book & Yelp2018 \\
\hline
\# User & 70,679 & 45,919 \\
\# Item & 24,915 & 45,538 \\
\# Interactions & 846,434 & 1,183,610 \\
\# Density & 4.8e-4 & 5.7e-4 \\
\hline
 & \multicolumn{2}{c}{Knowledge Graph} \\
\hline
\# Entities & 29,714 & 47,472 \\
\# Relations & 39 & 42 \\
\# Triplets & 686,516 & 869,603 \\
\hline
 & \multicolumn{2}{c}{Multimodal Features} \\
 \hline
\# Visual Features & 23817 & 41287 \\
\# Text Features & 24915 & 43620 \\
\hline
\end{tabular}
\end{table}

\subsubsection{Evaluation Metrics}
For a fair comparison, we adopt the same full sorting strategy as \cite{mu2022learning,zhang2021mining,wang2019kgat}. Specifically, for each user in the test set, we consider their uninteracted items as negative samples, and then output the preference scores of all items for the user in the test set. We evaluate performance using three widely adopted evaluation protocols: Recall@K, NDCG@K and Precision@K. By default, we set K to 20 and report the average metrics among all users in the test set.

\subsubsection{Baselines}
We evaluate the performance of SLIF-MR against three categories of baseline methods: general collaborative filtering, multimodal recommendation, and knowledge graph-based methods.
\begin{itemize}
\item {\textbf{BPR-MF}} \cite{rendle2012bpr} is a basic factorization model that relies on pairwise ranking loss for recommendation.
\item {\textbf{LightGCN}} \cite{he2020lightgcn} simplifies vanilla GCN by removing feature transformation and nonlinear activation modules.
\item {\textbf{NGCF}} \cite{wang2019neural} enhances user and item representations by utilizing high-order connectivities through embedding propagation on the user-item bipartite graph.
\item {\textbf{BM3}} \cite{zhou2023bootstrap} generates contrastive views by applying simple dropout augmentation to the embedded representations of users and items, thereby achieving self-supervised learning in multi-modal recommendations.
\item {\textbf{FREEDOM}} \cite{zhou2023tale} improves upon the LATTICE \cite{zhang2021mining} model by freezing the item-item graph, and designs a degree-sensitive edge pruning method for denoising the user-item interaction graph.
\item {\textbf{TMFUN}} \cite{zhou2023attention} optimizes item representations through an attention-guided multi-step fusion strategy and contrastive learning.
\item {\textbf{DiffMM}} \cite{jiang2024diffmm} proposes a multi-modal diffusion generative model that constructs a user-item graph incorporating multi-modal information through a step-by-step noise injection and denoising process, thereby enhancing the modeling of user preferences.
\item {\textbf{KGAT}} \cite{wang2019kgat} uses an attention mechanism to perform recursive information propagation on the knowledge graph, thereby explicitly capturing high-order connectivities.
\item {\textbf{KGCL}} \cite{yang2022knowledge} utilizes additional supervision signals from the knowledge graph to guide cross-view contrastive learning, thereby obtaining more robust knowledge-aware item representations.
\item {\textbf{KGIN}} \cite{wang2021learning} designs a new information aggregation scheme to recursively integrate relational sequence information from long-range connections, refining user intent information and encoding it into user and item representations.
\item {\textbf{KACL}} \cite{wang2023knowledge} constructs user-item interaction view and knowledge graph view, and adopts a knowledge-adaptive contrastive learning approach to extract shared information between the two views.
\item {\textbf{KGRec}} \cite{yang2023knowledge} leverages an attention mechanism to generate rational scores for triplets and conducts knowledge graph reconstruction and cross-view contrastive learning based on the scores.
\item {\textbf{KRDN}} \cite{zhu2023knowledge} introduces an adaptive knowledge refining strategy and a contrastive denoising mechanism; the former can distill high-quality KG triplets for aggregation, while the latter is responsible for pruning noisy implicit feedback.
\item {\textbf{DiffKG}} \cite{jiang2024diffkg} introduces a generative diffusion model to learn high-quality structural representations from the progressive perturbation and reconstruction of the knowledge graph, and effectively aligns user interests with knowledge semantics through collaborative-aware graph convolution and contrastive learning mechanisms
\end{itemize}

\subsubsection{Parameter Settings} 
The proposed SILF-MR model is implemented in Pytorch with an NVIDIA RTX 4090 GPU for accelerated computation. For a fair comparison, we fix the ID embedding size to 64 and the batch size to 4096 for all models. For baseline models, all other hyperparameters are configured according to their original settings. Model parameters are initialized using the Xavier initializer, and optimization is performed using the Adam optimizer. We conduct a grid search to tune hyperparameters, including the learning rate (selected from $\{10^{-4}, 10^{-3}, 10^{-2}\}$), the GCN layers $L$ (selected from $\{1, 2, 3, 4, 5\}$), the inter-modal consistency loss ratio $\beta$ (selected from $\{0.1, 0.3, 0.5, 0.7, 0.9\}$),  and the intra-modal consistency loss ratio $\gamma$ (selected from $\{10^{-5}, 10^{-4}, 10^{-3}, 10^{-2}, 10^{-1}\}$).

\begin{table*}[htbp]
\centering
\caption{Overall performance comparison. The best performance is highlighted in bold and the second to best is highlighted by underlines. All improvements are significant with $p-$value $\leq$ 0.05.}
\resizebox{\textwidth}{!}{%
\begin{tabular}{llcccccccccccccccc}
\toprule
\textbf{Dataset} & \textbf{Metric} & MF-BPR & NGCF & LightGCN & BM3 & FREEDOM & TMFUN & DiffMM & KGAT & KGCL & KGIN & KACL & KGRec & KRDN & DiffKG & \textbf{Ours} & \textbf{\%Imp}\\
\midrule
\multirow{3}{*}{Amazon-Book} 
& Recall@20     & 0.1147 & 0.1174 & 0.1248 & 0.1406 & 0.1473 & 0.1547 & 0.1767 & 0.1162 & 0.1502 & 0.1653 & 0.1568 & 0.1218 & \underline{0.1766} & 0.1727 & \textbf{0.1891} & \textbf{7.08} \\
& NDCG@20       & 0.0621 & 0.0611 & 0.0668 & 0.0735 & 0.0789 & 0.0826 & 0.0799 & 0.0588 & 0.0792 & 0.0888 & 0.0843 & 0.0636 & \underline{0.0965} & 0.0948 & \textbf{0.1024} & \textbf{6.11} \\
& Precision@20  & 0.0122 & 0.0124 & 0.0132 & 0.0149 & 0.0156 & 0.0165 & 0.0130 & 0.0124 & 0.0159 & 0.0177 & 0.0164 & 0.0136 & \underline{0.0188} & 0.0179 & \textbf{0.0204} & \textbf{8.51} \\
\midrule
\multirow{3}{*}{Yelp2018} 
& Recall@20     & 0.0590 & 0.0619 & 0.0615 & 0.0670 & 0.0717 & 0.0722 & 0.0817 & 0.0620 & 0.0753 & 0.0709 & 0.0666 & 0.0642 & \underline{0.0834} & 0.0822 & \textbf{0.0860} & \textbf{3.12} \\
& NDCG@20       & 0.0382 & 0.0400 & 0.0395 & 0.0432 & 0.0467 & 0.0468 & 0.0460 & 0.0401 & 0.0492 & 0.0461 & 0.0435 & 0.0416 & \underline{0.0545} & 0.0542 & \textbf{0.0565} & \textbf{3.67} \\
& Precision@20  & 0.0146 & 0.0152 & 0.0151 & 0.0164 & 0.0177 & 0.0178 & 0.0137 & 0.0153 & 0.0185 & 0.0177 & 0.0165 & 0.0159 & \underline{0.0205} & 0.0201 & \textbf{0.0212} & \textbf{3.41} \\
\bottomrule
\end{tabular}
}
\end{table*}

\subsection{Performance Comparison (RQ1)}
\subsubsection{Overall Performance Comparison}
Table II presents the performance comparison results, with the notation $\%Imp$ representing the percentage improvement over the best-performing baseline. From the experimental findings, we make the following key observations:

\begin{itemize}
\item The SLIF-MR model outperforms all other models across all datasets. This superior performance can be attributed to two primary factors: (i) The integration of item-item correlations as additional structural information within the heterogeneous graphs in a self-loop manner, which allows for more effective incorporation of underlying structural data. This leads to better representations of both users and items. (ii) SLIF-MR introduces an innovative heterogeneous auxiliary information fusion strategy that reduces the semantic gap across different information sources by applying consistency loss.
\item Multimodal and knowledge graph-based models generally surpass traditional collaborative filtering models. This advantage arises from the extensive item attribute data provided by multimodal features and the rich semantic relationships embedded within knowledge graph, both of which collectively help address the cold-start problem.
\item Among multimodal baselines, TMFUN achieves the best performance. This success is due to its multi-step fusion strategy, which effectively captures the intrinsic relationships between multimodal data and the user-item interaction graph.
\item Among the knowledge graph-based models, KRDN delivers the best performance. Its success is attributed to its use of contrastive learning for denoising, while simultaneously leveraging downstream supervisory signals to prune irrelevant triples, thereby enhancing recommendation accuracy.
\end{itemize}

\begin{figure}[!t]
  \centering
  % 先插入 yelp 图，使其在左边（反向排布）
  \subfloat[\normalfont{Amazon-Book}\label{BookAgg.pdf}]{%
    \includegraphics[width=0.49\linewidth]{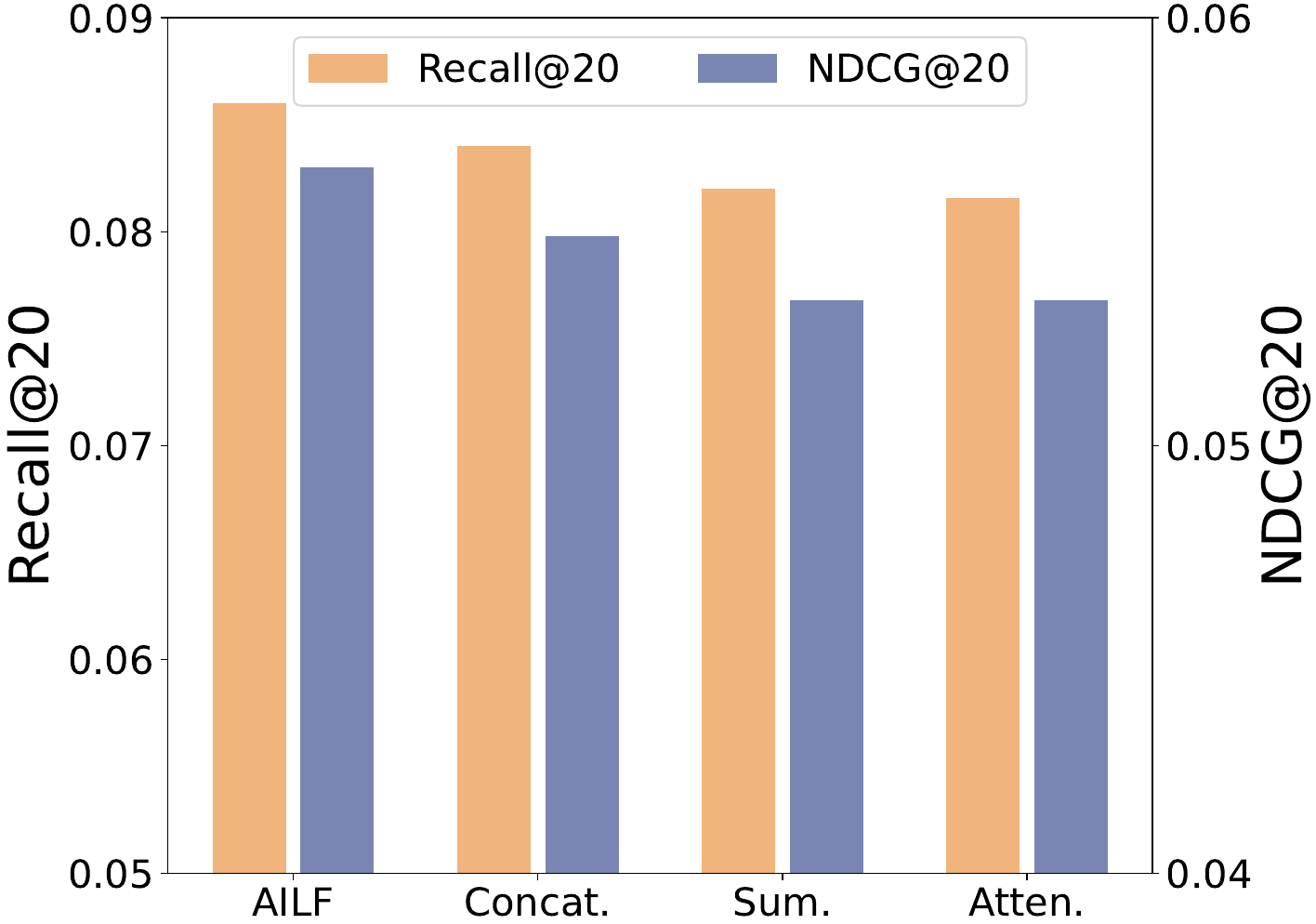}}%
  \hfill
  \subfloat[\normalfont{Yelp2018}\label{YelpAgg.pdf}]{%
    \includegraphics[width=0.49\linewidth]{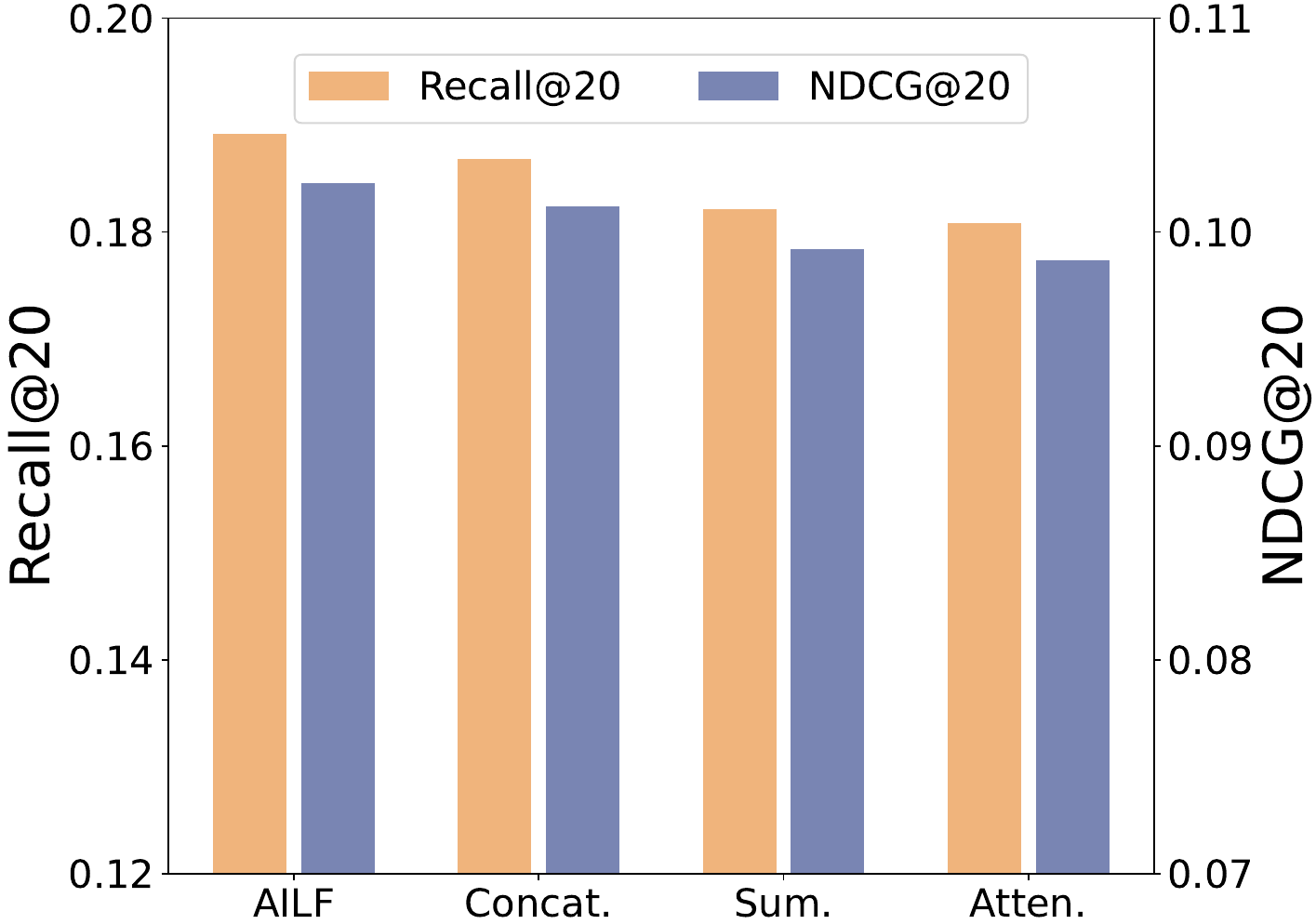}}%
  \caption{Model performance using different aggregation methods.}
  \label{fig:aggregation}
\end{figure}

\subsubsection{Aggregation Methods Comparison}
To assess the effectiveness of the proposed AILF method for aggregating heterogeneous representations, we compare it with three other commonly used aggregation techniques: concatenation \cite{wang2019kgat}, summation \cite{zhang2021mining}, and an attention-based aggregator \cite{zhou2023attention}. The experimental results, shown in figure \ref{fig:aggregation}, indicate that the AILF method consistently outperforms the other methods across key evaluation metrics on two datasets. This outcome highlights the significance of adaptively learning item-specific weight scores during aggregation, which allows for a more refined integration of heterogeneous representations.

\subsection{Ablation Study (RQ2)}
In this subsection, we examine the individual impact of the knowledge graph, multimodal information, their combination, the self-loop graph enhancement, and the consistency learning method.
\begin{itemize}
\item  w/o KG: We remove the knowledge graph from SLIF-MR and rely only on the multimodal item features as auxiliary information for recommendation.
\item  w/o MM: We eliminate the multimodal features from SLIF-MR and use only the knowledge graph of items as auxiliary information.
\item  w/o KG \& MM: We remove both the knowledge graph and the multimodal features of items from SLIF-MR.
\item  w/o SGE: We disable the self-loop heterogeneous graph enhancement module and use only the item feature graph to model potential item correlations.
\item  w/o CL: We remove the consistency loss and directly use the weighted aggregated heterogeneous item representations as the final item representations.
\end{itemize}

\begin{table}[htbp]
\centering
\caption{Impact of key components on SLIF-MR's performance.}
\begin{tabular}{llccc}
\toprule
\textbf{Dataset} & \textbf{Variants} & Recall@20 & NDCG@20 & Precision@20\\
\midrule
\multirow{5}{*}{Book} 
& w/o KG        & 0.1816 & 0.0998 & 0.0196\\
& w/o MM        & 0.1811 & 0.0990 & 0.0196\\
& w/o KG\&MM  & 0.1789 & 0.0977 & 0.0193\\
& w/o SGE       & 0.1851 & 0.1008 & 0.0201\\
& w/o CL        & 0.1817 & 0.0989 & 0.0196\\
& SLIF-MR        & 0.1891 & 0.1024 & 0.0204\\
\midrule
\multirow{5}{*}{Yelp2018} 
& w/o KG  & 0.0833 & 0.0546 & 0.0205\\
& w/o MM  & 0.0827 & 0.0540 & 0.0204\\
& w/o KG\&MM  & 0.0808 & 0.0529 & 0.0200\\
& w/o SGE  & 0.0818 & 0.0533 & 0.0201\\
& w/o CL  & 0.0835 & 0.0546 & 0.0207\\
& SLIF-MR        & 0.0860 & 0.0565 & 0.0212\\
\bottomrule
\end{tabular}
\end{table}

The results of the ablation study are presented in Table III. From these results, we draw the following key observations:
\begin{itemize}
    \item The complete SLIF-MR model consistently achieves the best performance, highlighting the significant contribution of each component to the overall effectiveness of the model.
    \item Removing either the knowledge graph or the multimodal features from SLIF-MR results in a decline in performance. The knowledge graph provides rich semantic relationships between entities, while multimodal features contribute detailed attribute-level information about items. Furthermore, excluding both the knowledge graph and multimodal features leads to a more substantial performance drop, as SLIF-MR effectively combines their complementary strengths through a well-structured fusion mechanism.
    \item The variant model without the SGE module performs worse than the full model, as the SGE module continuously refines the heterogeneous graph structures in a self-loop manner, gradually aligning them with true user preferences and item correlations.
    \item Omitting the consistency loss leads to a notable performance decline. This occurs because the inter-modal consistency loss bridges the semantic gap between heterogeneous data by aligning item representations across modalities. In addition, the intra-modal consistency loss addresses the problem of semantic distortion within individual modalities, preventing misalignment of similar items.
\end{itemize}

\begin{figure}[!t]
  \centering
  % 第一行两张图
  \subfloat[\normalfont{Varied $L$}\label{L}]{%
    \includegraphics[width=0.49\linewidth]{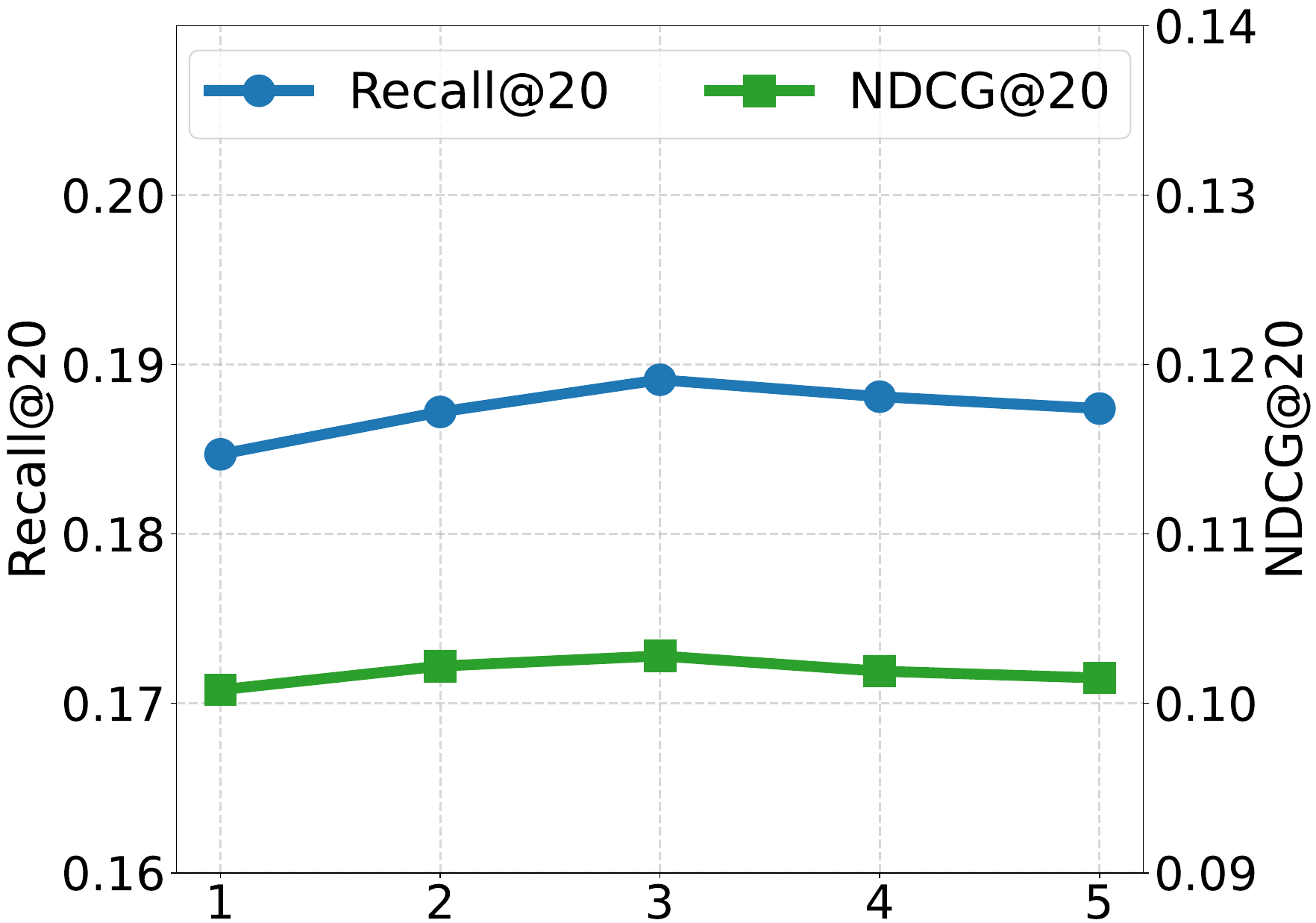}}
  \hfill
  \subfloat[\normalfont{Varied $\beta$}\label{beta}]{%
    \includegraphics[width=0.49\linewidth]{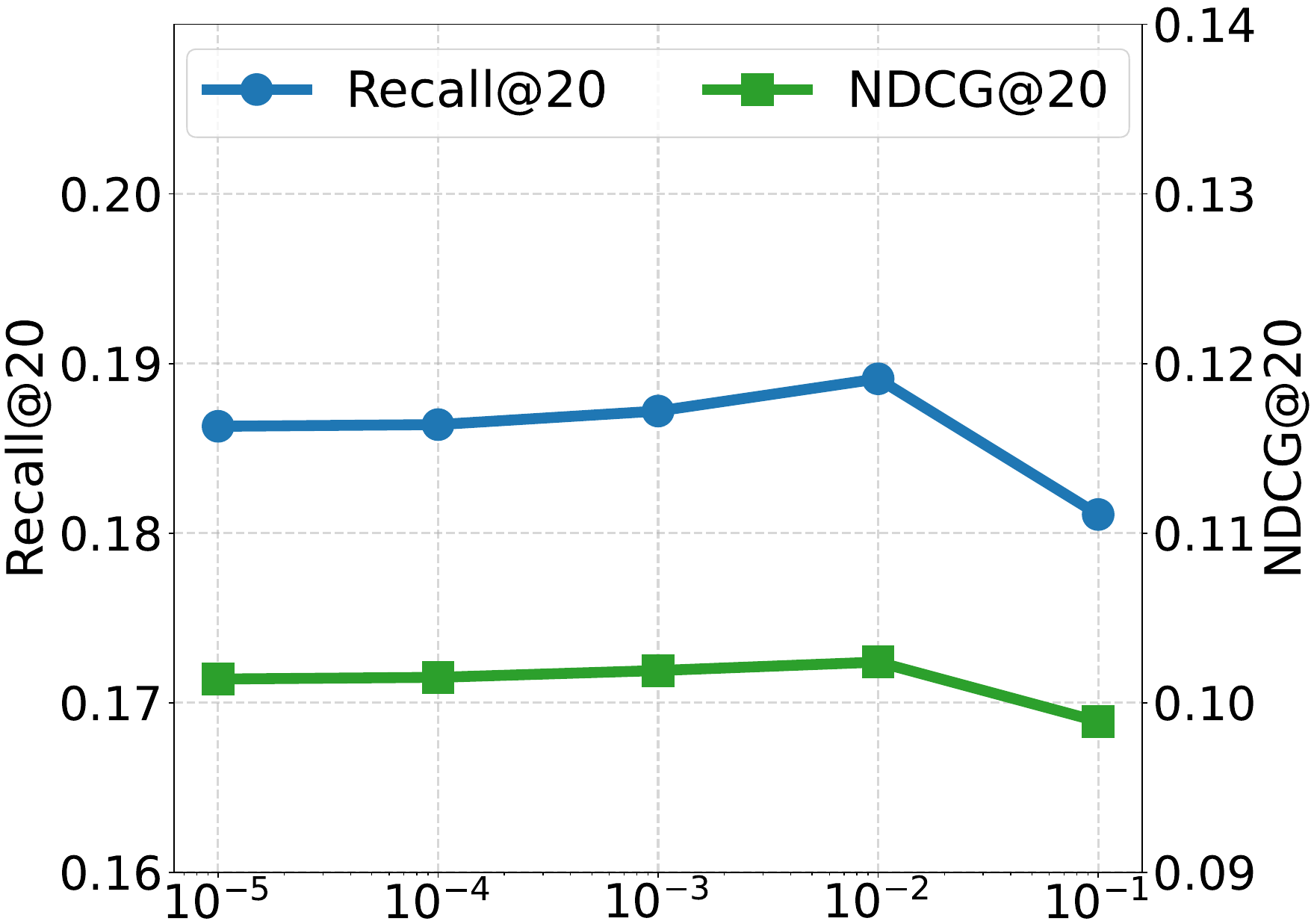}}
  \vspace{1.6mm} % <-- 缩小上下行的间距
  % 第二行两张图
  \subfloat[\normalfont{Varied $\gamma$}\label{gamma}]{%
    \includegraphics[width=0.49\linewidth]{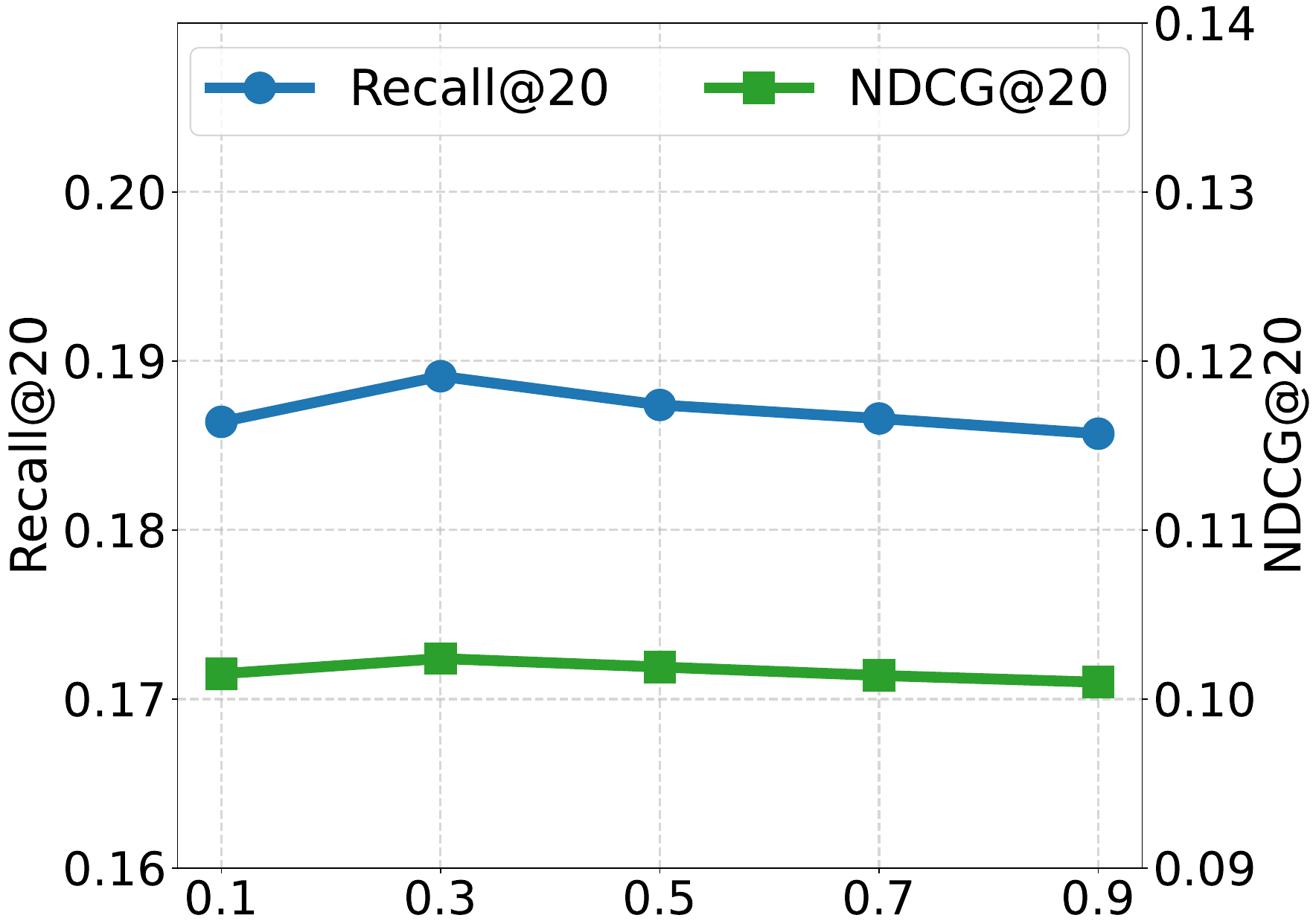}}
  \hfill
  \subfloat[\normalfont{Varied $k$}\label{topk}]{%
    \includegraphics[width=0.49\linewidth]{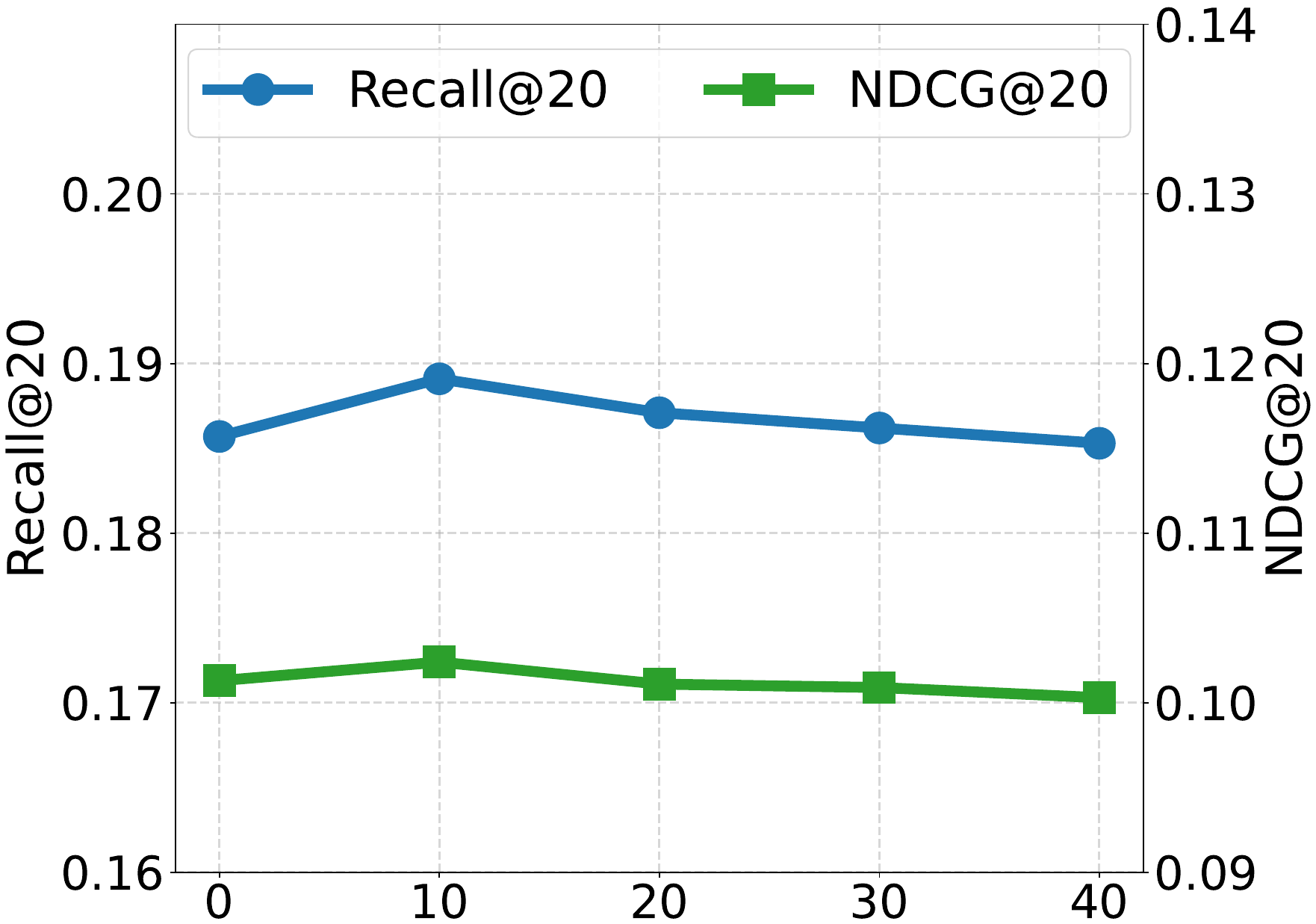}}
  \caption{Hyper-parameter sensitivity analysis on Amazon-Book.}
  \label{fig:hyperparameter}
\end{figure}

\subsection{Hyper-parameters Analysis (RQ3)}
As shown in Figure \ref{fig:hyperparameter}, we analyze the impact of several key hyperparameters on model performance. Due to space constraints, we present the results only for the Amazon-Book dataset, as the trends observed for the Yelp dataset are similar.
\begin{itemize}
    \item Figure \ref{fig:hyperparameter} (a) illustrates the performance trend as the number of GCN layers, $L$, increases. Initially, the performance improves, reaches a peak, and then fluctuates downward. This behavior occurs because deeper GCN layers allow nodes to aggregate information from a broader neighborhood range. However, as $L$ continues to grow, the overlap of neighborhoods among different nodes increases, leading to over-smoothing. This results in more similar representations across nodes, which hinders the model’s ability to capture personalized user preferences.
    \item The impact of the cross-modal consistency loss coefficient is illustrated in Figure \ref{fig:hyperparameter} (b). As $\beta$ increases, the distances between heterogeneous item representations in the embedding space gradually decrease, leading to a corresponding reduction in the semantic differences among heterogeneous information. However, as this process continues, the heterogeneous item representations tend to homogenize, resulting in the loss of unique information inherent in different types of data.
    \item Figure \ref{fig:hyperparameter} (c) shows that model performance improves initially and then declines as the intra-modal consistency loss coefficient, $\gamma$, increases. A higher value of $\gamma$ promotes a more uniform distribution of item representations within the same modality, preventing "abnormal proximity" among dissimilar items due to cross-modal consistency learning. However, excessively high $\gamma$ values lead to overly distinct item representations, impairing the model’s ability to capture true item correlations.
    \item As the value of $N$ increases, the proposed SGE module enhances the modeling of implicit correlations between items, thereby better uncovering user preferences. However, as $N$ continues to grow, items with minimal similarity may be incorrectly linked, introducing noise into the model.
\end{itemize}

\begin{table}[htbp]
\centering
\caption{Self-loop complexity analysis.}
\begin{tabular}{llccccc}
\toprule
\textbf{Dataset} & \textbf{Metric} & \multicolumn{5}{c}{\textbf{Epoch Interval}} \\
\cmidrule(lr){3-7}
& & \textbf{1} & \textbf{5} & \textbf{10} & \textbf{15} & \textbf{20} \\
\midrule
\multirow{2}{*}{Book} 
& Recall@20 & 0.1891 & 0.1871 & 0.1855 & 0.1851 & 0.1844 \\
& NDCG@20   & 0.1024 & 0.1013 & 0.1003 & 0.1001 & 0.0993 \\
\midrule
\multirow{2}{*}{Yelp2018} 
& Recall@20 & 0.0860 & 0.0856 & 0.0848 & 0.0845 & 0.0840 \\
& NDCG@20   & 0.0565 & 0.0560 & 0.0556 & 0.0554 & 0.0552 \\
\bottomrule
\end{tabular}
\end{table}

\subsection{Robust Study (RQ4)}
To comprehensively evaluate the robustness of the proposed SLIF-MR model under various noise conditions, we design a series of controlled experiments. Following the dataset partitioning strategy described in Subsection 4.1, we maintain a fixed test set while injecting different types of noise into the training and validation sets. Specifically, for interaction noise, we randomly replace a certain proportion of original user-item interactions with noisy interactions. For knowledge graph noise, we randomly select a percentage of triples and replace their tail entities. Additionally, to simulate modality absence, we randomly remove varying proportions of image and text features from the item representations.

\begin{figure*}[!t]
  \centering
  % 第一行整体作为一个子图 (a)
  \subfloat[Amazon-Book: recommendation accuracy comparison with different noise types (from left to right: interaction noise, knowledge noise and modality noise)]{%
    \begin{minipage}[b]{0.99\textwidth}
      \centering
      \includegraphics[width=0.32\textwidth]{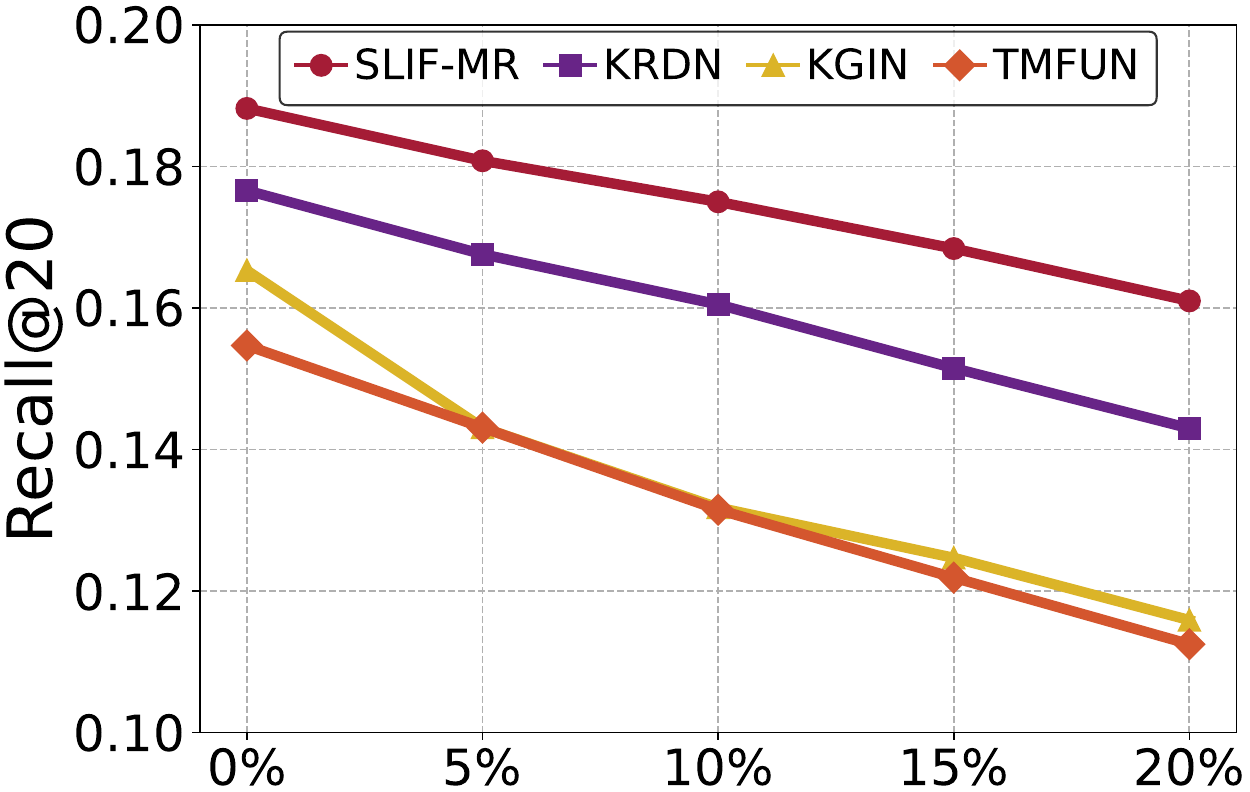}
      \hfill
      \includegraphics[width=0.32\textwidth]{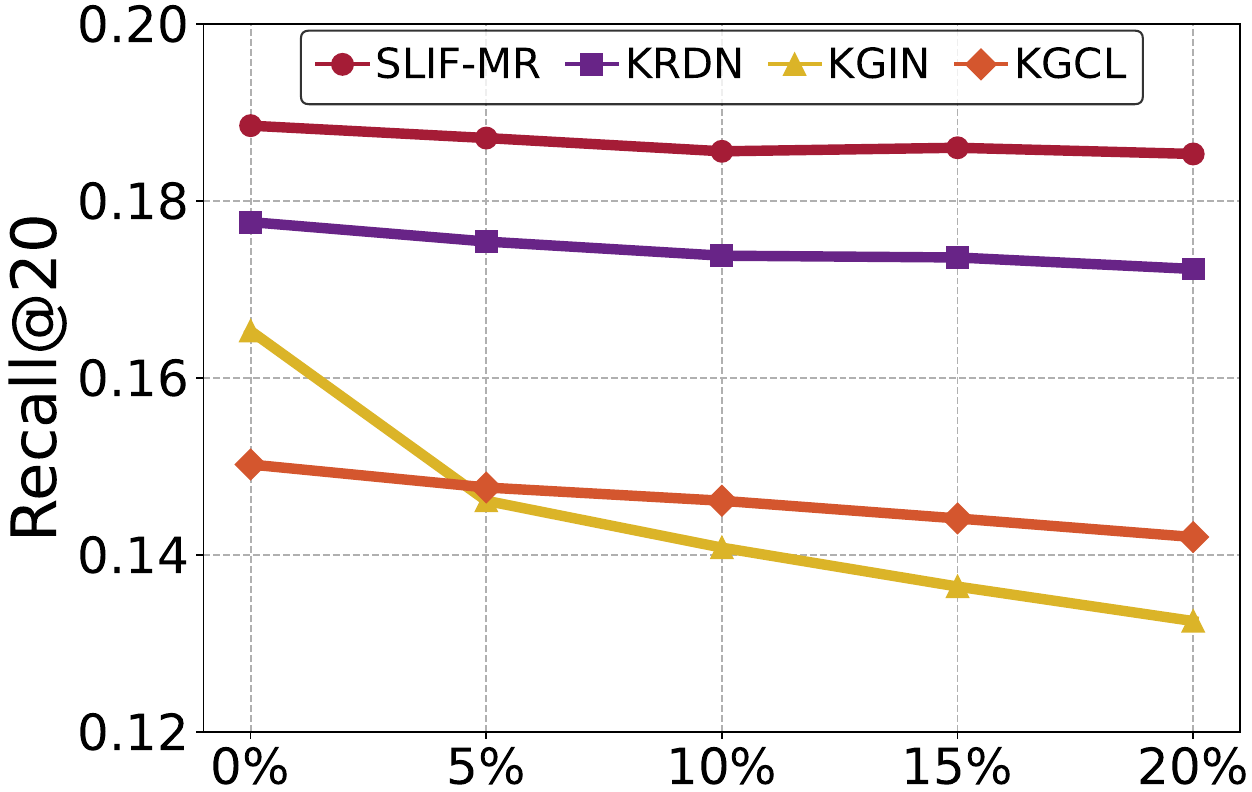}
      \hfill
      \includegraphics[width=0.32\textwidth]{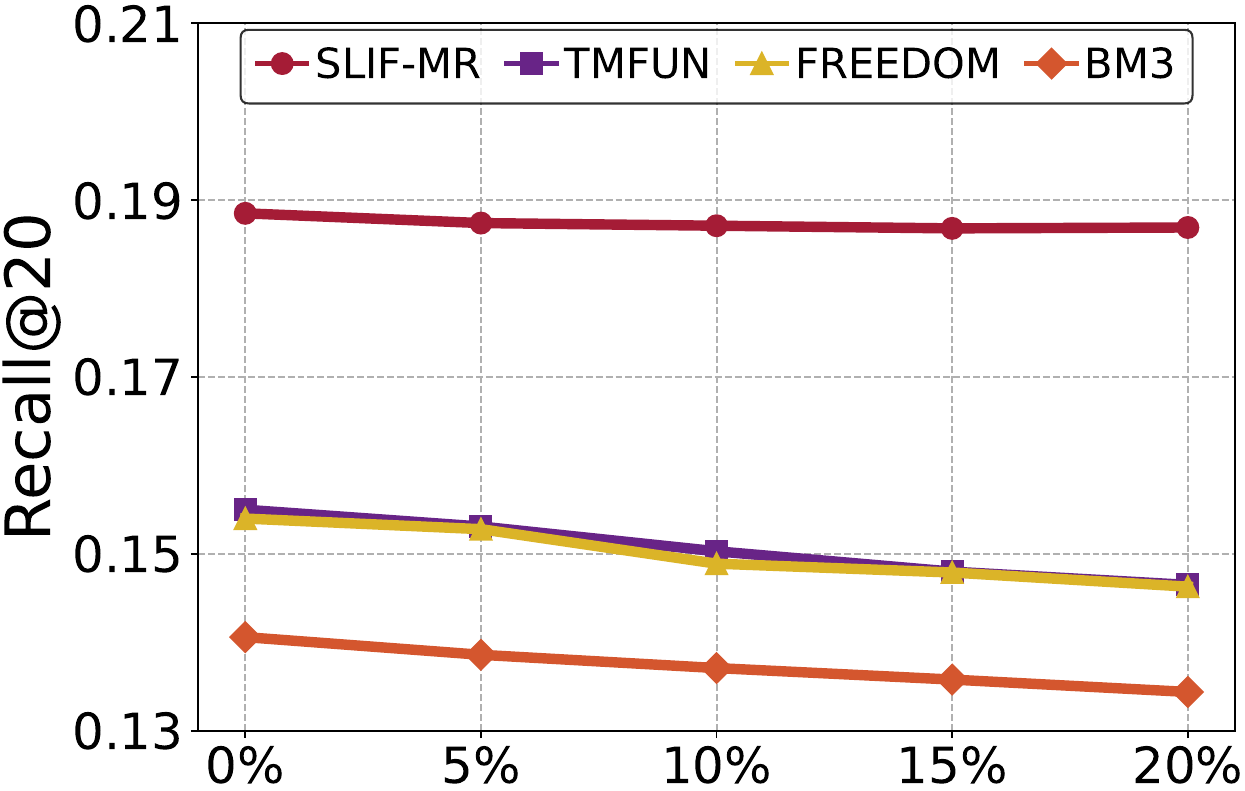}
    \end{minipage}
  }\\
  % 第二行整体作为一个子图 (b)
  \subfloat[Yelp2018: recommendation accuracy comparison with different noise types (from left to right: interaction noise, knowledge noise and modality noise)]{%
    \begin{minipage}[b]{0.99\textwidth}
      \centering
      \includegraphics[width=0.32\textwidth]{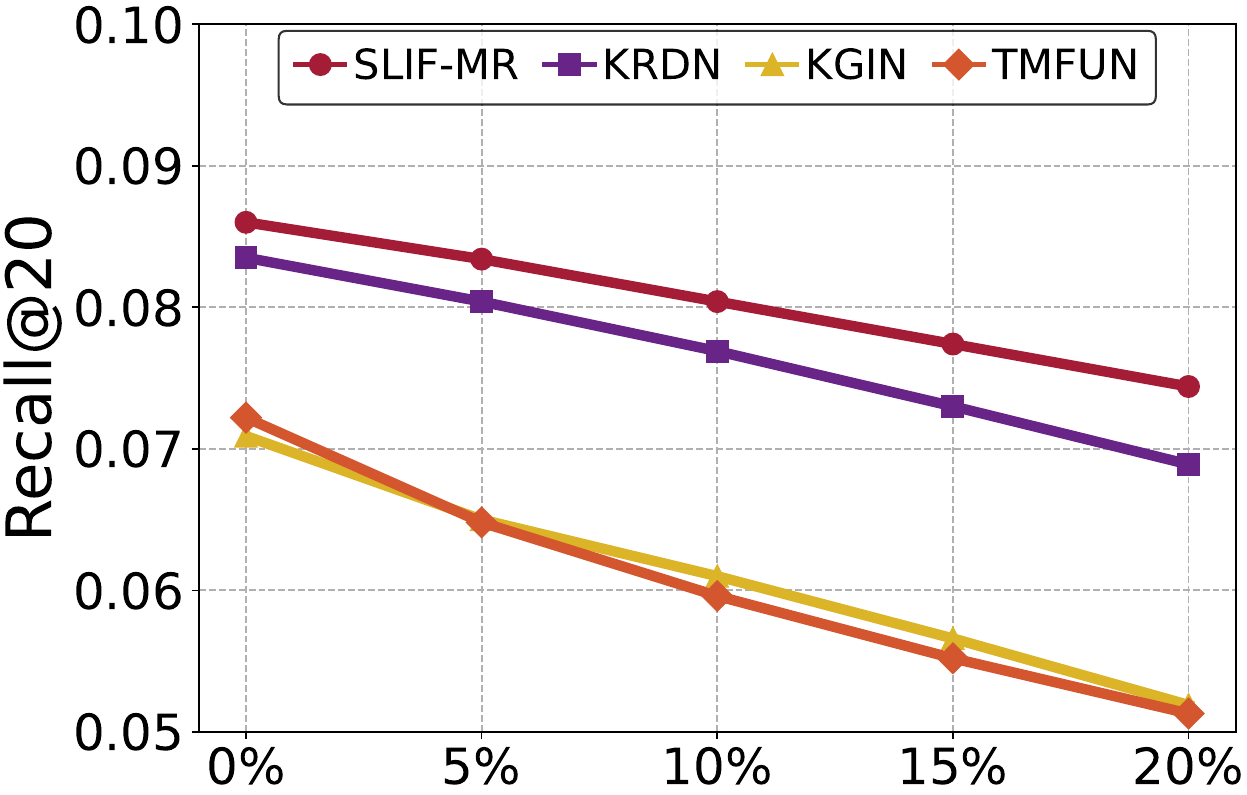}
      \hfill
      \includegraphics[width=0.32\textwidth]{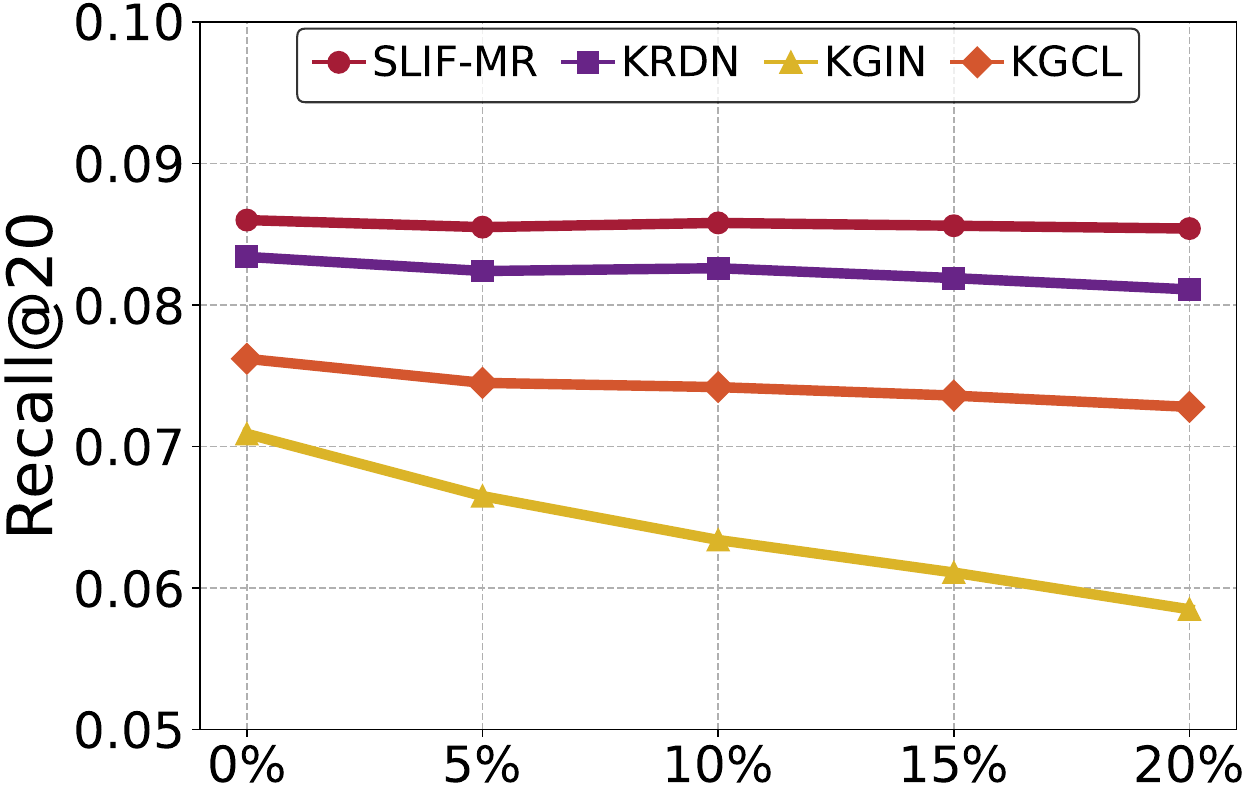}
      \hfill
      \includegraphics[width=0.32\textwidth]{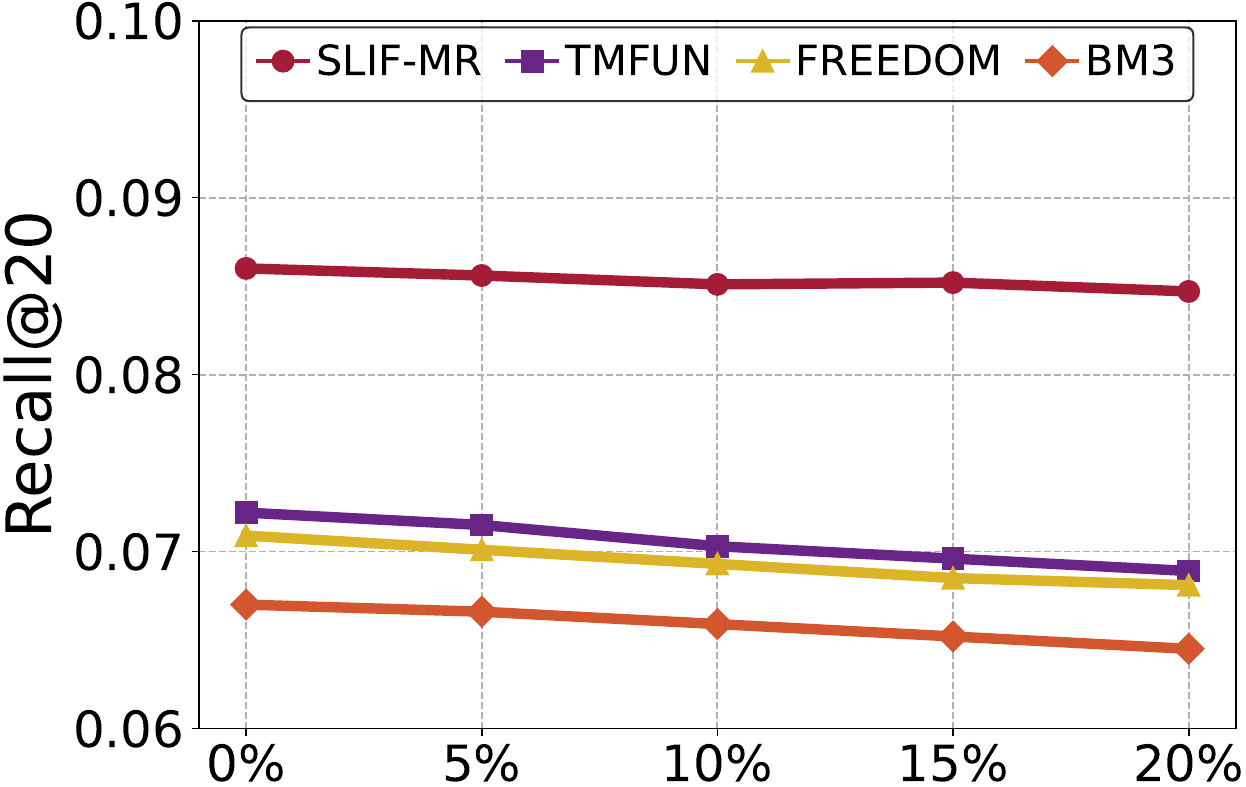}
    \end{minipage}
  }
  \caption{Performance comparison of model robustness.}
  \label{fig:noise}
\end{figure*}
The robustness results are illustrated in Figure \ref{fig:noise}. As the proportion of interaction noise increases, the performance of all baseline models deteriorates markedly. In particular, in Amazon-Book dataset, SLIF-MR exhibits only a 15.3\% reduction in Recall@20 when the noise level reaches 20\%, whereas the best-performing baseline suffers a performance decline of 19.6\%. This demonstrates that SLIF-MR's self-loop graph enhancement mechanism effectively optimizes the interaction graph structure, progressively aligning it with genuine user preferences and thereby significantly mitigating the adverse effects of interaction noise.

Moreover, SLIF-MR consistently outperforms baselines under knowledge graph corruption and modality absence conditions. For example, in Yelp datset, under 20\% knowledge graph noise, SLIF-MR achieves a performance drop that is 1.7\% smaller compared to the strongest baseline. Similarly, in scenarios with 20\% missing modality information, SLIF-MR retains 3.1\% higher Recall@20 than competing models. These results highlight that the semantic consistency learning strategy effectively bridges the modality and knowledge heterogeneity, enabling SLIF-MR to robustly integrate complementary information and maintain high recommendation accuracy under noisy environments.

\subsection{Self-loop Complexity Analysis (RQ5)}
To analyze the computational efficiency of the self-loop enhancement mechanism, we examine the impact of varying update epoch intervals on model performance. As reported in Table IV, increasing the epoch interval between self-loop graph updates from 1 to 20 results in a gradual yet consistent decline in both Recall@20 and NDCG@20 across the Amazon-Book and Yelp2018 datasets. This performance degradation reflects a weakening of the model's ability to iteratively refine item-item correlations in alignment with evolving user preferences. However, SLIF-MR maintains relatively stable and competitive results even under sparse update conditions, highlighting the robustness of the proposed self-loop design and its ability to balance recommendation quality with computational cost.

\section{Conclusion}
In this paper, we propose SLIF-MR, a novel recommendation framework that fully exploits heterogeneous auxiliary information through self-loop iterative fusion. By dynamically constructing the item-item correlation graph based on feedback item representations and injecting it into the heterogeneous graph structures, SLIF-MR progressively aligns graph topology with user preferences. Moreover, the incorporation of a semantic consistency learning mechanism effectively mitigates the semantic gap among heterogeneous information, resulting in unified and robust item representations. Extensive experiments on two real-world datasets demonstrate that SLIF-MR significantly outperforms state-of-the-art baselines in terms of accuracy and robustness. In future work, we aim to investigate collaborative denoising strategies for heterogeneous data sources and extend our framework to incorporate additional modalities such as video, audio, and social relationship graphs.

\bibliographystyle{IEEEtran}
\bibliography{main}
\end{document}